 \documentclass[twocolumn,floatfix,tighten]{aastex62}

\usepackage{amsmath}
\usepackage{amssymb}
\usepackage{graphicx}
\usepackage{bm}

\definecolor{todo}{RGB}{200,0,0}

\received{}
\revised{}
\accepted{}
\submitjournal{ApJ}

\shortauthors{DiGiorgio Zanger et al.}
\shorttitle{Bisymmetric Modes in SDSS-IV/MaNGA}

\begin{document}

\title{The Strength of Bisymmetric Modes in SDSS-IV/MaNGA Barred Galaxy Kinematics}

\correspondingauthor{Brian DiGiorgio Zanger}
\email{bdz100@juniata.edu}

\author[0000-0003-3344-7776]{Brian DiGiorgio Zanger}
\affiliation{Department of Physics, Juniata College, 1700 Moore St., Huntingdon, PA 16652, USA}
\affiliation{Department of Physics and Astronomy, Colby College, 4000 Mayflower Hill, Waterville, ME 04901, USA}
\affiliation{Department of Astronomy and Astrophysics, University of California, Santa Cruz, 1156 High St., Santa Cruz, CA 95064, USA}

\author[0000-0003-1809-6920]{Kyle B. Westfall}
\affiliation{University of California Observatories, University of California, Santa Cruz, 1156 High St., Santa Cruz, CA 95064, USA}

\author{Kevin Bundy}
\affiliation{Department of Astronomy and Astrophysics, University of California, Santa Cruz, 1156 High St., Santa Cruz, CA 95064, USA}
\affiliation{University of California Observatories, University of California, Santa Cruz, 1156 High St., Santa Cruz, CA 95064, USA}

\author{Niv Drory}
\affiliation{ McDonald Observatory, University of Texas at Austin, 1 University Station, Austin, TX 78712, USA}

\author{Matthew A. Bershady}
\affiliation{Department of Astronomy, University of Wisconsin-Madison, 475N. Charter St., Madison WI 53703, USA}

\author{Stephanie Campbell}
\affiliation{School of Maths, Statistics and Physics, Newcastle University, Newcastle upon Tyne, NE1 7RU}
\affiliation{School of Physics and Astronomy, University of St. Andrews, North Haugh, St. Andrews KY16 9SS, UK}

\author{Anne-Marie Weijmans}
\affiliation{School of Physics and Astronomy, University of St. Andrews, North Haugh, St. Andrews KY16 9SS, UK}

\author{Karen L. Masters}
\affil{Departments of Physics and Astronomy, Haverford College, 370 Lancaster Avenue, Haverford, PA 19041, USA}

\author{David Stark}
\affil{Space Telescope Science Institute, 3700 San Martin Drive, Baltimore, MD 21218, USA}
\affil{Department of Physics and Astronomy, Haverford College, 370 Lancaster Avenue, Haverford, PA 19041, USA}

\author{David Law}
\affiliation{Space Telescope Science Institute, 3700 San Martin Drive, Baltimore, MD 21218, USA}

\begin{abstract}


The SDSS-IV/MaNGA Survey data provide an unprecedented opportunity to study the internal motions of galaxies and, in particular, represent the largest sample of barred galaxy kinematic maps obtained to date. We present results from Nirvana, our non-axisymmetric kinematic modeling code built with a physically-motivated Bayesian forward modeling approach, which decomposes MaNGA velocity fields into first- and second-order radial and tangential rotational modes  in a generalized and minimally-supervised fashion. We use Nirvana to produce models and rotation curves for 1263 unique barred MaNGA galaxies and a matched unbarred control sample We present our modeling approach, tests of its efficacy, and validation against existing visual bar classifications. Nirvana finds elevated non-circular motions in galaxies identified as bars in imaging, and bar position angles that agree well with visual measurements. The Nirvana-MaNGA barred and control samples provide a new opportunity for studying the influence of non-axisymmetric internal disk kinematics in a large statistical sample. 

\end{abstract}



\section{Introduction}
\label{sec:intro}

Galactic bars are smooth linear bisymmetric morphological features in the central regions of disk galaxies \citep{binney08}. A large fraction of disk galaxies in the local Universe have bars, including the Milky Way \citep{blitz91}, with more massive, redder galaxies having larger bar fractions \citep{nair10, masters11}. Barred galaxies have been observed out to $z>2$ \citep{guo22} and been observed to be long-lived in the local Universe \citep{gadotti15}, though studies disagree on whether bar fraction decreases with redshift or remains steady, with some evidence that dynamical disturbances and large gas inflows can disrupt existing bars \citep{gadotti15, kraljic12, melvin14, cameron10, sheth08, elmegreen04}. 

Bars are inherently dynamical structures stemming from perturbations in a galaxy's gravitational potential that lead to destabilizing resonances in stellar orbits \citep{athanassoula02} and the redistribution of angular momentum throughout the disk \citep{kormendy04}. Spontaneous bar formation has been observed in galaxy evolution simulations ranging from relatively simple models of galactic potentials \citep[e.g.][]{toomre81}, to low-resolution n-body simulations \citep[e.g.][]{sellwood93}, to modern hydrodynamical simulations \citep[e.g.][]{rosas22}. Bars can also form due to changes in galactic potential from major mergers or tidal disruptions \citep{bi22} and can evolve over the course of a galaxy's lifetime.

The dynamical structure of bars can be seen through the motions of material within the galaxy. Bars channel interstellar gas radially along their leading edge \citep{regan97}, with gas flowing both inwards and outwards \citep{fragkoudi16}. This radial motion also redistribute stellar populations within bars, flattening population gradients within the bar as compared to the surrounding disk \citep{fraser19}. These motions may play a part in the early quenching of star formation in barred galaxies \citep{fraser20}. These structures can also be studied using the Tremaine-Wineberg method \citep{tremaine84}, allowing for the determination of bar pattern speed and corotation radius in spatially-resolved spectroscopy of samples of barred galaxies and further insight into the potential of the dark matter halo, gas fraction, and star formation history \citep[e.g.][]{geron23,garma20,garma22,cuomo21}.


Conventional single-geometry velocity field models \citep[e.g.][]{andersen2013} describe ordered circular rotation in disk galaxies using simple analytic models to derive global kinematic parameters like inclination, position angle, and asymptotic speed. However, these methods are limited in their application to only galaxies that can be reasonably modeled as a single dynamical system, so for non-axisymmetric galaxies with bars, warps, or other disruptions, a more flexible formalism is needed. Tilted ring models \citep[e.g.][]{begeman87, begeman89, jozsa07, oh18} forego a global kinematic model and instead describe the kinematics using a series of discrete concentric rings with independent kinematic parameters, and  \citet{stark18} describes position angle variation continuously as a function of radius for non-axisymmetric galaxies using the Radon transform. Kinemetry \citep{krajnovic06} uses the techniques of surface photometry to perform harmonic decomposition of the higher-order spatial modes present in 2D velocity fields of irregularly-rotating galaxies. However, without additional assumptions about galaxy structure and rotation curves, the models resulting from these methods do not have an explicit astrophysical interpretation.

\textit{Velfit} (\citealp{spekkens07, sellwood10}, \citealp[later \textit{DiskFit,}][]{sellwood15}) instead proposes a single cohesive model for a galaxy's disk properties. Based on harmonic models from \citet{schoenmaker97}, the \textit{Velfit} model has global values for inclination and position angle, instead accounting for kinematic distortions with added modes on top of the usual first-order (i.e. completing one sinusoidal velocity oscillation per revolution) tangential velocity of a circularly-rotating disk. They use only physically-motivated terms in their model, restricting it to fitting either first-order radial term that accounts for sloshing or a combination of second-order radial and tangential terms that are meant to represent bisymmetric motions within bars. These models have had success in describing non-circular motions in radio observations of cold gas rotation in nearby galaxies \citep[e.g.][]{bisaria22, garma22, holmes15} and have been re-developed using a Bayesian framework called \texttt{XookSuut} \citep{lopezcoba21}. However, all of these models use piece-wise nonparametric rotation curve models, which are more flexible for describing unanticipated motions but provide less physical insight. 

In this paper, we build on these earlier kinematic models of non-circular motions to create Nirvana,\footnote{Nonaxisymmetric Irregular Rotational Velocity ANAlysis, available at \url{https://github.com/briandigiorgio/NIRVANA}.} a flexible code for modeling bisymmetric motions in barred galaxies. We develop our model using a Bayesian forward modeling framework {similar to \textsc{2DBAT} \citep{oh18} but} with added constraints within the prior and tuning of the likelihood function that are adjusted to produce more robust, physically-viable results than are possible with simple least-squares optimizers. Additional features include point-spread function (PSF) convolution, dispersion fitting, and surface brightness weighting to make the model more easily applied to velocity fields where the size of the PSF not small relative to the galaxy {as compared to existing models like DiskFit}, allowing for analysis in regimes that were ill-suited to previous methods. We investigate the biases present in the model using mock data to calibrate results.

We apply the Nirvana model to a sample of barred galaxies from the SDSS-IV MaNGA\citep{bundy15}. Using bar designations from volunteer classifications of MaNGA galaxy morphology from GalaxyZoo: 3D \citep[GZ:3D;][]{masters21}, we attempt to fit the stellar and gas-phase velocity fields of all barred MaNGA galaxies and model their non-circular motions with Nirvana, as well as a population-matched sample of unbarred galaxies that we use as a control, generating corresponding samples of velocity field models. We find elevated levels of bisymmetric motion in the barred sample as compared to the unbarred control, and we find that galaxies with elevated bisymmetric velocity terms generally match GZ:3D closely in bar position angle. 

This paper is structured as follows: Section \ref{sec:data} summarizes the galaxy kinematic data we use and how we prepare it for modeling, as well as the assembly of the samples of barred and unbarred galaxies. Section \ref{sec:bisym} describes our velocity model and PSF convolution methods. Section \ref{sec:fitting} describes Nirvana's fitting algorithm, including the prior and likelihood functions in the Bayesian model. Section \ref{sec:assessments} discusses our evaluations of the model's effectiveness when compared to real and mock data. Section \ref{sec:summary} provides a summary of our work and presents directions for future study. 

\section{MaNGA Data}
\label{sec:data}

\subsection{MaNGA: Mapping Nearby Galaxies at Apache Point Observatory} \label{sec:manga}


This paper utilizes data and data products from the Sloan Digital Sky Survey IV \citep[SDSS-IV;][]{york00, blanton17} and the Mapping Nearby Galaxies at Apache Point Observatory survey \citep[MaNGA][]{bundy15}. MaNGA uses integral field spectroscopy to collect spatially-resolved spectra for $\sim$10,000 galaxies using the BOSS spectrographs on the 2.5 m telescope at Apache Point Observatory \citep{gunn06}. Spectral observations have a resolution of $R \sim 2000$ over a range of $3600$ \AA $< \lambda < 10300 $\AA with variable exposure time to achieve the desired signal-to-noise ratio (SNR) of 10 in the $g$-band \citep{bundy15}. Fibers are grouped into hexagonal bundles of 19 to 127 fibers that are 12" to 32" in diameter \citep{drory15}. Flux calibration and sky subtraction are applied to the observed spectra using simultaneous observations of standard stars and sky within the same field \citep{yan16}. The median full-width half-maximum (FWHM) of the point-spread function (PSF) for MaNGA data cubes is 2.5", which roughly corresponds to kiloparsec scales at the targeted redshifts ($z < 0.15$). Observations are dithered and interpolated onto a 0.5" grid of spaxels.

The MaNGA sample is selected to be uniform over i-band absolute magnitude and is divided into two subsamples: the Primary+ sample ($\sim$2/3 of the total sample) that contains galaxies with spectral coverage out to $\sim$1.5 effective radii ($R_e$), and the Secondary sample ($\sim$1/3 of the total sample) where observations extend out to $\sim$2.5 $R_e$ \citep{wake17}. Raw spectroscopic observations are reduced by the MaNGA Data Reduction Pipeline \citep[DRP;][]{law16}, and data products such as velocity measurements are derived with the Data Analysis Pipeline \citep{westfall19,belfiore19}. All data in this paper are from the seventeenth SDSS data release \citep[DR17;][]{dr17}, which represents the final data release of the MaNGA survey and contains MaNGA observations and data products from 10,010 unique galaxies. All photometric data in this paper is from the NASA-Sloan Atlas \citep[NSA;][]{blanton11}, which uses imaging from SDSS-I, II, and III and assumes $H_0 = 100$ km/s/Mpc.

In this paper, we utilize the hybrid binning scheme data products from the DAP, which uses slightly different methods for creating stellar- and gas-phase line-of-sight velocity measurements. For the stellar kinematics, spaxels are Voronoi binned \citep{cappellari03} to a threshold $g$-band-weighted SNR of at least 10.  These bins are then deconstructed such that the gas kinematics are determined on a spaxel-by-spaxel basis. Both velocity fields are calculated by simultaneously fitting all emission/absorption lines, meaning that all ionized gas tracers are assumed to have the same velocity. For this reason, for the remainder of the paper, when we discuss velocity fields derived from observations of nebular emission, we refer to them as ``gas-phase" velocity fields rather than velocity fields associated with a particular emission line. However, each emission line is fit independently for surface brightness and velocity dispersion, so we use the H-alpha values for these quantities when working with gas-phase velocity data.

\subsection{Data Processing} \label{sec:clipping}

Though the MaNGA DAP masks many imperfections in the maps it extracts from the datacubes, there are still outliers in the data that inhibit our ability to produce a successful fit. 

Specifically, the DAP also sometimes produces velocity measurements for individual spaxels that differ greatly from the neighboring spaxels due to systematic errors caused by low SNR \citep{westfall19, belfiore19}. To identify these spurious velocity measurements, we convolve a kernel to blur the kinematic data that is equivalent to the reported PSF, smearing the data over a scale that should correspond to the observational differences in the data. We then mask any spaxels where the magnitude of the discrepancy between the velocity and dispersion maps and their blurred  counterparts, since any spaxels that differ too greatly from their neighbors must be nonphysical. Through experimentation, we determined any spaxels with discrepancies of more than 50 km/s are likely erroneous, so they are masked.  


We then mask out any spaxels that have a surface brightness flux of less than $3 \times 10^{-19}$ ergs/s/cm$^2$ per spaxel in the $H\alpha$ flux map or an $H\alpha$ amplitude-to-noise ratio (ANR) of less than 5 for gas velocity fields, or $3 \times 10^{-19}$ ergs/s/cm$^2/\rm{\AA}$ per spaxel in the stellar flux map for stellar velocity fields. These values were experimentally determined to best remove low-quality velocity measurements on the outskirts of galaxies.

Finally, we attempt to remove any regions of the velocity field that do not appear to be part of the same rotating system as the rest of the galaxy. Many MaNGA IFUs contain foreground/background sources or merging companions that have distinct velocity fields from the main target, so it would be inappropriate to fit a single rotating disk to the data. To mask these, we perform a preliminary fit to the kinematics using an axisymmetric model using a hyperbolic tangent rotation curve and subtract the model from the data to obtain a map of the residuals.  If the data are well represented by this model, the residuals should be randomly distributed along a Gaussian distribution according to the Central Limit Theorem, and any deviations from Gaussianity represent possible signatures of asymmetry that we may want to mask. In order to preserve the genuine bisymmetric features we are attempting to model, we mask only the spaxels that differ from the mean of the residuals by more than 10 standard deviations, a value we experimentally determined removes unwanted companions but still preserves real bisymmetric features. After masking these spaxels, we again fit the axisymmetric model and remove the outliers in the residuals, repeating the process until the number of masked spaxels stabilizes.

\begin{figure}
    \centering
    \includegraphics[width=\columnwidth]{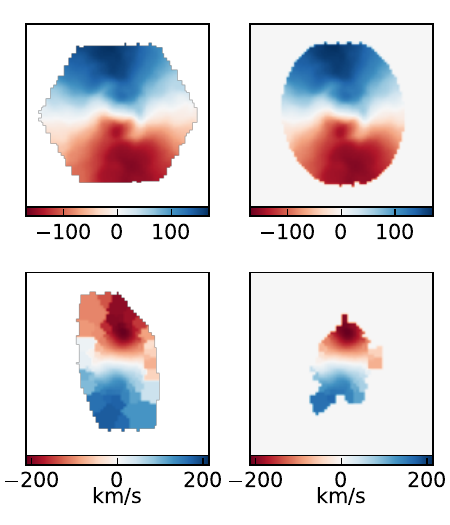}
    \caption{Example velocity fields before (left) and after (right) the masking process for a high SNR gas-phase velocity field (MaNGA plate-IFU number 8078-12703, top) and a stellar-phase velocity field with relatively low SNR in the outskirts (11750-9101, bottom). Spaxels with outlying velocities, or low SNR, flux, or ANR are masked. The method is detailed in Section \ref{sec:clipping}.}
    \label{masking}
\end{figure}

If, at the end of this process, the galaxy is left with only 20\% or less of its original number of spaxels unmasked, the velocity field is considered to be unsuitable for velocity field fitting and it is not fit. Less than 1\% of sample galaxies fall below this threshold, and the median fraction of masked spaxels is less than 10\%. Two illustrations of the the masking process are shown in Figure \ref{masking}, with one high SNR gas-phase velocity field (top) and one relatively low SNR stellar-phase velocity field (bottom). 

Our rotation curve models are piece-wise linear functions defined on a set of concentric elliptical rings. While a set of parametric rotation curve functions would be more computationally efficient and easier to physically interpret, there is currently little evidence available to construct such rotation curve functions for bisymmetric modes, pushing us to instead use flexible piece-wise functions to describe the motions as closely as possible, leaving the construction of a parametric bisymmetric model for future work. 

To construct the radius of each ring, we determine the position of the minor axis and inclination of the galaxy our preliminary axisymmetric model (see above) and transform the spaxel/bin coordinates into in-plane elliptical coordinates
. We then subdivide these coordinates into concentric rings using the method described further in Section \ref{sec:bisym}. If more than 75\% of the spaxels in a given elliptical annulus are masked, all spaxels are discarded and the relevant ring is removed. This prevents a small number of spaxels from having an undue influence on the model, particularly in galactic outskirts. Any galaxies with 2 or fewer elliptical rings are discarded for having insufficient spatial resolution. 

\subsection{Sample}
\label{sec:sample}

Our goal is to assess the ability of Nirvana to accurately model and quantify bisymmetric distortions in the velocity fields of MaNGA galaxies. To this end, we define two galaxy samples, one of barred galaxies where we expect prominent bisymmetric kinematic distortions, and a second matched control population of galaxies that do not appear to be barred (see Section \ref{sec:control}). To create these samples, we use the existing Galaxy Zoo: 3D catalog \citep[GZ:3D;][]{masters21}, a crowd-sourced project for identifying morphological features in SDSS images of MaNGA galaxies. Volunteers drew regions on images of all MaNGA galaxies from the SDSS-I/II survey \citep{gunn98, york00} to indicate which morphological feature each pixel belonged to, yielding vote counts for each pixel that we can use to determine which galaxies have bars as well as the shape of the bar. We chose this catalog over others because it already provides information on bar position and shape within the galaxy, allowing us to more easily compare our models to existing imaging.

We define a pixel as being part of the bar if more than 20\% of volunteers designated it as such, and we define a galaxy as ``barred" if it has more than at least one spaxel that is part of a bar, the methodology recommended by \citet{krishnarao20} and \citet{masters21}. GZ:3D provides us not only with a binary classification of barred versus unbarred galaxies but also with more detailed spatial information that we will compare to our kinematic modeling results. In the MaNGA sample, there are 1593 such galaxies representing 14.1\% of the total sample. Since MaNGA provides both stellar and gas velocity maps, we model both using Nirvana, but fit the two tracers independently. 

Major mergers can greatly disrupt the internal kinematics of disk galaxies, so we also remove any galaxies that are obviously undergoing a merger. GZ:3D has volunteers mark the centers of any galaxies that are in the image of the target galaxy and the surrounding area, so we remove any galaxies where the average number of centers marked by volunteers was greater than 1.5, indicating that a majority of volunteers found more than one center, a threshold we determined by visual inspection. We find a total of 98 mergers in our original list of barred galaxies and remove them from our final sample to reduce extra sources of non-circular motion.

After these cuts, Nirvana produces velocity field models for 973 stellar velocity fields (66.6\% of the initial sample) and 1012 gas-phase velocity fields (69.3\%). 722 galaxies (49.4\%) have both stellar and gas velocity fits, and 1263 unique galaxies have either a stellar or gas-phase velocity field model. These sets of successfully fit galaxies represents our final Nirvana-MaNGA sample of barred galaxies that we will work with for the remainder of this paper.

\begin{figure}
    \centering
    \includegraphics[width=\columnwidth]{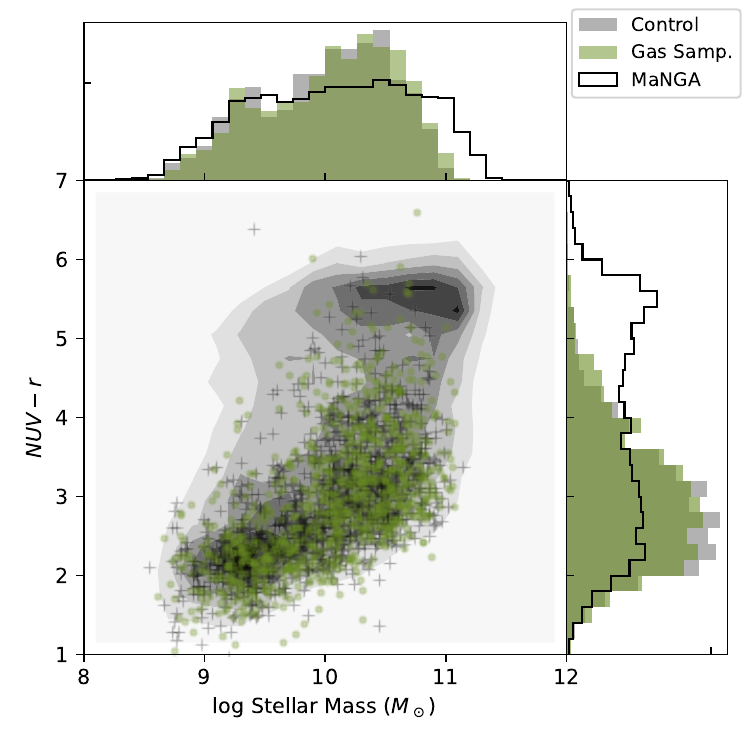}
    \caption{Stellar masses and colors of the Nirvana gas-phase sample of barred galaxies (green circles), the population-matched control sample (gray pluses), and the MaNGA sample as a whole (contours). The sample galaxies lie almost entirely within the ``blue cloud," with only a small number having green or red colors, and there is a greater fraction of high-mass blue galaxies than in the overall MaNGA sample. The control sample of unbarred galaxies is demographically extremely close to the Nirvana sample by virtue of the matching process.}
    \label{cmd}
\end{figure}

The cuts in our data processing tend to bias the Nirvana-MaNGA sample away from redder galaxies because of their lower gas-phase emission flux, resulting in a sample of galaxies that fall almost entirely within the ``blue cloud" of galaxies on the color-magnitude diagram. As shown in Figure \ref{cmd}, the majority of the sample lies between $10^9 - 10^{11} M_\odot$, as described by the elliptical Petrosian photometry data given in the NSA \citep{blanton11}. The sample is almost entirely blue, as measured by the NSA elliptical Petrosian $NUV - r$, with only a few galaxies in the green valley and red sequence. There are peaks in the mass distribution around $10^{9.3}$ and $10^{10.4}$. The first peak corresponds to a mass range with large representation in the overall MaNGA sample of blue galaxies, and the second indicates a bias towards larger blue galaxies overall within the Nirvana sample. If bar-driven secular evolution does indeed lead to quenching \citep{gadotti15}, then this may indicate that our galaxies have relatively recently formed bars, but further study of stellar populations in the bars is necessary to confirm this.

\subsection{Control Sample}
\label{sec:control}

To isolate the effect of galactic bars on our main sample, we construct a sample of unbarred galaxies to serve as a control. Such a sample will allow us to compare the strength of the bisymmetric distortions measured by Nirvana to our main sample, where the bisymmetric distortions are expected to be more significant.  This control should therefore resemble the population of galaxies in our main sample, such that we can effectively isolate the effect of the bars.  To build the control sample, we match each barred galaxy in the final sample to a galaxy with similar NSA elliptical Petrosian stellar mass, color, axis ratio ($b/a$), and half-light radius ($R_{50}$) using linear sum assignment \citep{llsum}, which produces a set of unique galaxy pairs with matched population parameters.

For each of the parameters listed, we normalize the range of the MaNGA population to fall roughly between 0 and 1.\footnote{Some ranges were chosen to capture the range of galaxies in the full MaNGA sample, so they may appear oversized when considering just our sample of barred galaxies. However, changing the bounds has only a small effect on the overall properties of the galaxies chosen for the control.} The end points of the normalized parameter distributions are as follows:
\begin{itemize}
    \item Color ($NUV-r$): 0 to 10.
    \item Log stellar mass: $10^8$ to $10^{12} M_\odot$.
    \item Half-light radius: 0 to 18 arcsec.
    \item Axis ratio: 0 to 1.
\end{itemize}
The median distance between galaxy pairs in the normalized parameter space 0.038, so the population statistics for the control sample are nearly identical to the barred sample, as seen in Figures \ref{cmd}.

\section{Bisymmetric Kinematic Model}
\label{sec:bisym}

To model non-circular motions in disk galaxies, we adopt a formalism based on \cite{spekkens07}. Our models use a cylindrical coordinate system, with the disk plan at $z=0$, projected on the sky.  To  map the rectilinear on-sky spaxel coordinates onto the projected galaxy coordinates, we use the following transformations:
\begin{eqnarray}
    r & = & \left[(x - x_c)^2 + (y - y_c)^2\right]^{1/2} \\
    \theta & = & \arctan\left(\frac{x \sin \phi - y \cos \phi}{\cos i\, (x \cos \phi + y \sin \phi)}\right),
\end{eqnarray}
for $x$ and $y$ center position $x_c$ and $y_c$ and on-sky position angle $\phi$, measured from N through E along the direction of the receding side of the major axis. 

We split the velocity field $V(r,\theta)$ into its radial and tangential components $V_r(r,\theta)$ and $V_t(r,\theta)$, additionally breaking each component down into its Fourier modes. \citet{spekkens07} show that some bisymmetric (second-order) terms are degenerate with a first-order radial term. Here, we neglect the first-order radial term, effectively assuming that most galaxies have no radial sloshing. 

We limit our model to only the primary rotation term (first-order tangential) and second-order terms to focus on the bisymmetric flows that are physically associated with bars, rather than higher-order modes that may describe local non-bisymmetric irregularities in velocity fields more exactly \citep[e.g.][]{krajnovic06}. However, \citet{spekkens07} note that sinusoidal models of order $m$ projected in an elliptical coordinate system are degenerate with models of order $m \pm 1$, so some third-order features are present in the models. We address the first-order degeneracies in Section \ref{sec:likelihood}. 

The resulting model is shown below:
\begin{eqnarray} \label{model}
    V(r, \theta) & = & V_{sys} + \sin i \, \bigg[V_t(r) \cos \theta \nonumber \\ & & - V_{2t}(r) \cos \big(2 (\theta - \phi_b)\big) \cos \theta \nonumber \\ & & - V_{2r}(r) \sin \big(2 (\theta - \phi_b)\big) \sin \theta \bigg].
\end{eqnarray}
The bisymmetric position angle $\phi_b$ is defined as the in-plane angular difference between the first- and second-order rotational terms. We also discretize the kinematic components, $V_t, V_{2t},$ and $V_{2r}$, using a piece-wise linear function with breakpoints at equally-spaced in-plane radii.  The breakpoint radii are set such that their separation is defined as half of the reconstructed FWHM of the MaNGA PSF along the minor axis of the galaxy, thus Nyquist sampling the changes in velocity along the position angle where they are most compressed. These breakpoint radii are linearly spaced along the minor axis until the edge of the MaNGA IFU is reached, as described  in Section \ref{sec:clipping}. Additional details regarding the construction of the kinematic models are addressed in Section \ref{sec:fitting}.  We note that the inner-most breakpoint of the functions is at $R=0$, and we force all velocity components to be 0 km/s at this position.

Nirvana also goes beyond previous works by simultaneously modeling the velocity dispersion of the input galaxy. In addition to providing a more complete kinematic understanding of the galaxy, the dispersion also helps to more accurately model the effects of beam smearing by incorporating both spatial and spectral smearing in the final velocity measurements. The increased fidelity and generality of our beam smearing also differentiates Nirvana from prior work \citep[e.g.][which was restricted to a PSF width was less than the width of the annular ring]{spekkens07}. Since velocity dispersion is a second-order moment, we assume that it is radially symmetric \citep{binney08}. Therefore, we do not need a complex model to decompose it like we do for the velocity, instead modeling it as a single piece-wise curve $\sigma(r)$ defined over the radius of the galaxy and projected in-plane. However, such simple axisymmetric models may be limited in their ability to describe galaxies that are not axisymmetric themselves and particularly because bars themselves do cause some local increases in velocity dispersion along the bar axis and at the ends of the bar \citep{du16}.

Once the intrinsic models for velocity and dispersion have been generated, they are convolved with the MaNGA PSF to include the effects of beam smearing, which can be directly compared with the observed data.  The convolutions performed are
\begin{eqnarray}
    I_{\rm mod} & = & I \ast P, \\
    V_{\rm mod} & = & \frac{(I\ V) \ast P}{I_{\rm mod}}, {\rm and} \\
    \sigma_{\rm mod} & = & \left[\frac{I (V^2 + \sigma^2) \ast P}{I_{\rm mod}} - V_{\rm mod}^2\right]^{1/2},
\end{eqnarray}
where $\ast$ is the convolution operator, $P$ is the on-sky point-spread function, and {the quantities $I$, $V$, and $\sigma$ are all the intrinsic properties of the galaxy along the line-of-sight, before convolution with the PSF, and $I_{mod}, V_{mod},$ and $\sigma_{mod}$ are their modeled counterparts.}  Note that a limitation of our model is that we \textit{do not} model the surface brightness distribution, $I$, directly \citep[cf.][]{2019MNRAS.485.4024V}.  Given the computation expense of the latter,\footnote{Ideally, we would replace this approximation with, e.g., narrow-band H$\alpha$ imaging that has higher spatial resolution than the kinematic data.} we instead substitute the \textit{observed} surface brightness distribution (i.e., we replace $I \sim I_{\rm obs}$) in the equations above.

\section{Fitting Algorithm}
\label{sec:fitting}

The core function of Nirvana is to represent the input galaxy using the model described above. To fit the above model to the data, we construct a Bayesian forward model. We choose this formalism rather than a least-squares optimizer like \cite{spekkens07} because of its ability to compensate for local minima in the likelihood, account for covariances between parameters, and utilize priors when navigating probability space. We specifically chose the Bayesian code \texttt{dynesty} \citep{dynesty}, a Python package implementing nested sampling \citep{skilling04, skilling06} utilizing multi-ellipsoid bounds \citep{feroz09}, due to its strengths in describing high-dimensional multi-modal likelihood spaces. By randomly sampling the parameter space, nested sampling is able to constrain the posterior probability distribution while not getting stuck in local minima like a least-squares optimizer or a walker-based approach like Markov Chain Monte Carlo may. {The advantages of nested sampling were similarly recognized by \citet{oh18} in their own high-dimensional multi-modal Bayesian velocity field code \textsc{2DBAT.}}

In this section, we describe the prior and likelihood functions used by Nirvana as well as the biases and constraints that led to their design, expanding upon earlier Bayesian velocity field models \citep{lopezcoba21} by utilizing specially-designed prior and likelihood terms to better calibrate output. An example of the results from running the model is given in Section \ref{sec:example}.

\subsection{Priors} \label{sec:priors}

\subsubsection{Position angles, velocities, and centers}

To keep the fitting process relatively galaxy-agnostic, we endeavored to keep the priors as uninformative as possible. We chose a uniform prior over all angles for position angle $\phi$ rather than setting a narrower prior probability distribution based on preliminary axisymmetric fits to allow for complicating factors such as irregular galaxy shapes or non-circular motions that could lead to significant biases in the axisymmetric position angle. Similarly, we use a uniform prior over all angles for the second order position angle $\phi_b$ since we do not have any information on the likely orientations of higher order components for any of the galaxies. 

Rather than attempting to construct an informed prior for the individual velocity components based on predicted rotation curve shapes, we instead attempt to be neutral and keep the model as free from parametric models as possible by using uniform priors over a reasonable velocity range. We allow the magnitudes of the individual in-plane velocity components $V_t, V_{2t}$, and $V_{2r}$ to vary between 0 and 400 km/s in each ring, with the center held fixed at 0 km/s. Similarly, the prior on velocity dispersion magnitude $\sigma$ is uniform over 0 to 300 km/s. 

We have found that axisymmetric fits are almost always capable of recovering the systemic velocity well, so we restrict the $V_{sys}$ to be within $\pm$60 km/s of the value returned by the preliminary fit. We also rely on the MaNGA IFU placement for the position of the center of the galaxy, restricting the galactic center to be within a 4$\arcsec$ square box surrounding the center of the MaNGA bundle. We determined the size of the bounding box by noticing that in preliminary runs, almost all galaxy models that had kinematic centers more than 2$\arcsec$ from the IFU center were fit incorrectly, and that the results from the fit were improved by restricting the position of the dynamical center. Essentially all isolated galaxies are centered in the MaNGA IFU, and galaxies with kinematic centers outside of this bounding box are almost always not isolated or are undergoing a merger, making them unsuitable for our modeling approach.

\subsubsection{Inclination}
\label{sec:incbias}

The most restrictive prior we have placed on the fitting algorithm is on the inclination, which we tie to the photometric inclination using a relatively tight Gaussian prior. We derive the photometric inclination $i_p$ of each galaxy from its elliptical Petrosian axis ratio $q$, as provided by the NASA-Sloan Atlas \citep{blanton11}. We convert this value to a photometric inclination as follows:
\begin{equation} \label{photinc}
    \cos ^2 i_p = \frac{q^2 - q_0^2}{1 - q_0^2},
\end{equation}
where $q_0$ is the intrinsic oblateness of the galaxy. We do not have any information on the value of $q_0$ for each individual galaxy since such information would require detailed dynamical modeling of each galaxy, though it tends to correlate with scale length in late-type galaxies \citep{bershady10}. However, previous studies \citep[e.g.][]{weijmans14, padilla08, lambas92} find that for rotating galaxies like disks and fast-rotating ellipticals, $q_0 \approx 0.20-0.25$, so we choose a nominal value of $q_0 = 0.2$ for all galaxies in our model, similar to the \citet{bershady10} estimate of 0.25. These inclinations are more reliable than kinematically-derived inclinations from axisymmetric fits that are sometimes adversely affected by kinematic asymmetries, but the imprecision in these estimates may still have effects on the derived photometric inclination depending on a specific galaxy's morphology.

\begin{figure}
    \centering
    \includegraphics[width=0.4\textwidth]{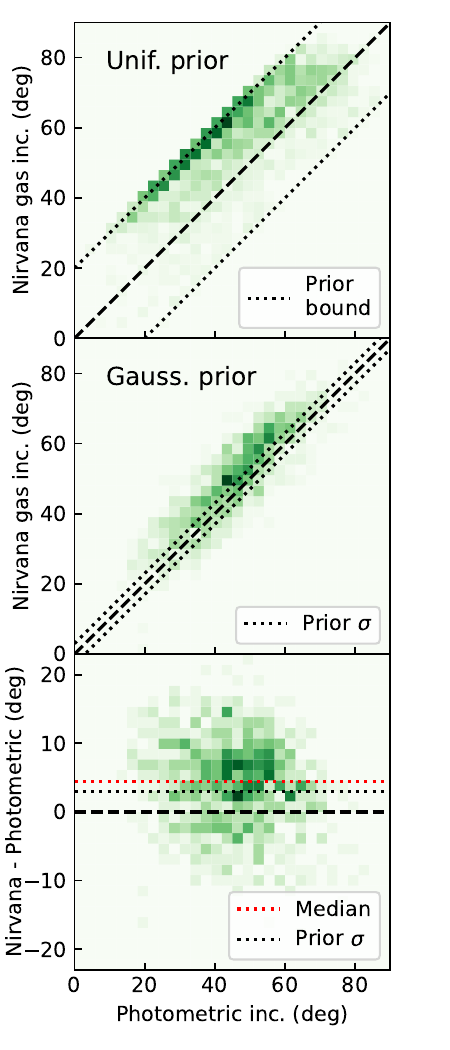}
    \caption{The effects of different inclination priors described in Section \ref{sec:priors} on the inclination recovered by Nirvana as compared to inclinations derived from photometry. Top: A uniform prior centered on the photometric inclination with a width of $\pm20^\circ$. Nirvana has a significant tendency to produce inclinations that are much too high, often running up against the prior bound (dotted line). Middle: A Gaussian prior centered on the photometric inclination with a standard deviation of $3^\circ$ (dotted line) produces a much better agreement with photometry while still allowing some freedom in the fit. Bottom: A comparison between the inclinations derived from photometry and the inclinations recovered by Nirvana in our sample of barred galaxies with a Gaussian prior. There is a systematic bias of $4-5^\circ$, which is in line with biases seen in other similar models.}
    \label{incpriors}
\end{figure}

We originally defined the prior as uniform distribution with bounds $\pm 20^\circ$ from $i_p$. {However, inherent degeneracies in the Nirvana model cause a strong tendency to fit galaxy inclinations that are significantly higher than either the input inclinations (in the case of mock galaxies) or the inclination derived from photometry (in the case of real data). These degeneracies stem from the similar appearances of the second-order terms and the velocity field residuals from an incorrect inclination, meaning that a model including both inclination and second-order velocity terms will experience this confusion.} While we do not expect perfect correspondence between the kinematic and photometric inclinations because they are tracing different components of the galaxy's structure, the systematic bias to high inclination indicated an underlying problem with the current state of the priors. As shown in the top panel of Figure \ref{incpriors}, most models were driven to the upper limit of this uniform prior.

{In lieu of a more complex exploration of this degeneracy (e.g. with a Fisher matrix analysis), we instead counteract this bias with a stricter inclination prior tied to photometric inclination measurements. We used a Gaussian prior centered on the photometric inclination and a $3^\circ$ standard deviation, which we determined in early testing to lessen the degeneracy}. This tighter constraint leads to much closer agreement with the photometric inclination, with the bias reduced to $4-5^\circ$, as seen in Figure \ref{incpriors}. A bias of this size is not much larger than that of existing axisymmetric models \citep[e.g.][]{andersen2013, digiorgio21},{ and \textsc{2DBAT} \citep{oh18} also experiences significant inclination biases in mock galaxies.} However, if the photometric and kinematic inclinations are indeed vastly different, e.g. in a galaxy with multiple kinematically-decoupled components, this prior is still flexible enough to allow Nirvana to fit the disk correctly.\footnote{Such misaligned structures are more common in early-type galaxies \citep{corsini14}, which are almost entirely absent from the Nirvana-MaNGA sample, so this situation is unlikely to be a major factor when Nirvana is applied to barred spirals.} The bottom panel of Figure \ref{incpriors} shows a comparison between the photometric inclinations of MaNGA galaxies calculated using Equation \ref{photinc} and the kinematic inclinations recovered by Nirvana. Due to the inherent degeneracy between inclination and rotational velocity, these stronger priors also have an effect on the recovered velocity profiles since elevated inclinations necessitate lower in-plane velocities for the same LOS velocity. The smaller model inclinations favored by the more restrictive prior brings velocity magnitudes back up to expected levels, rather than being biased low for the previous prior.

\subsection{Likelihood}
\label{sec:likelihood}

The Nirvana likelihood function is based primarily on a standard Gaussian likelihood. At each iteration of the fitting process, we generate a velocity field model according to the steps outlined in Section \ref{sec:bisym} using the latest parameter guesses. We then compute a $\chi^2$ value between the original data and the model, weighting each spaxel by its velocity variance $\sigma_v^2$ as reported by the MaNGA DAP with an extra error term of 5 km/s added in quadrature, an extra term to provide an error floor in cases where the DAP produces erroneously low errors. Summing over all unmasked spatial elements, we obtain one value for the whole galaxy:
\begin{equation} \label{chisq}
    \chi_v^2 = \sum \frac{(V-V_{\rm mod})^2}{\sigma_v^2}.
\end{equation}

We calculate separate $\chi^2$ likelihoods for the velocity and dispersion data, substituting in the square of the physical velocity dispersion 
\begin{equation}
\sigma^2 = \sigma_{\rm obs}^2 - \sigma_{\rm corr}^2,
\end{equation}
where $\sigma_{\rm obs}$ is the velocity dispersion reported by the MaNGA DAP and $\sigma_{\rm corr}$ is an instrumental correction \citep{westfall19}.\footnote{The correction factor for the gas kinematics is the instrumental resolution at the best-fitting line wavelength; for the stellar kinematics, it is a correction that accounts for the difference in spectral resolution between the MaNGA spectra and the stellar templates used to measure the kinematics.}  Using the reported errors on dispersion $\sigma_\sigma$ plus an error floor of 5 km/s added in quadrature, the resulting chi-squared term is as follows:
\begin{equation}
    \chi_\sigma^2 = \sum_{\rm elem.} \frac{(\sigma_-\sigma_{\rm mod})^2}{\sigma_\sigma^2}.
\end{equation}
In our modeling of the velocity dispersions, we do not account for systematic errors in the measurements caused by inaccuracies in the MaNGA line-spread function, low signal-to-noise, or effects of truncating the error distribution to consider only corrected values with $\sigma^2 > 0$ \citep[][Chattopadhyay et al., submitted]{2021AJ....161...52L}.  These issues will be considered in more detail in future work.  Here, we note that these biases have relatively little influence on our velocity field fits, the primary concern of this paper.

The resulting chi-squared terms are then added together as part of the final likelihood. In addition to these chi-square terms, we include specific penalty functions that mitigate biases and unphysical results discovered while testing our approach. Although these penalties of arbitrary form and come at the expense of the objectivity of the modeling procedure, they provide more robust final results. We describe each penalty, $P_1$ and $P_2$ in the following two subsections. The final likelihood function $L$ is represented by:
\begin{equation}
    \log L = -\chi_v^2 - \chi_\sigma^2 - P_{1,v} - P_{1,\sigma} - P_2.
\end{equation}

\subsubsection{Smoothing penalty}

To incentivize the model to produce smoother radial profiles, we impose a penalty if the second derivative of the rotation curve shape is high for any of the components. We approximate the second derivative by taking the difference between the kinematic components in each concentric ring and the mean of the values of the same component in neighboring ring, equivalent to convolution with kernel [1,-2,1] across the piece-wise rotation curve. The smoothing penalty $P_1$ is the sum of the second derivative for all ring, scaled by the magnitude of the velocity component in that ring and weighted by a coefficient $w_1$:
\begin{equation} \label{smoothing}
    P_1 = w_1 \sum_{i}^{N_{rings}} \frac{V_i - (1/2)(V_{i-1} + V_{i+1})}{V_i}.
\end{equation}
This penalty is applied for all velocity components as well as the velocity dispersion, and the resulting penalty is subtracted from the log likelihood. We determined that a weight of $w_{1,v} = 10$ for velocity components and $w_{1,\sigma}=1$ for velocity dispersion. These values result in rotation curves adequately describe spatially-coherent differences in velocity as a function of radius in mock galaxy trials while also moderating sharp (and often unphysical) discontinuities in the shapes of the velocity profiles. 

\subsubsection{Second-order velocity penalty}

Testing of mock galaxies shows a notable covariance in the posteriors of the inclination and the second-order radial component of the velocity $V_{2r}$. The velocity field residuals for an improper inclination have similar patterns to the effects of $V_{2r}$, resulting in Nirvana sometimes preferring to return inclinations that were too high and then counteract the residuals from that mistake with elevated $V_{2r}$ values. 

\begin{figure}
    \centering
    \includegraphics[width=.5 \textwidth]{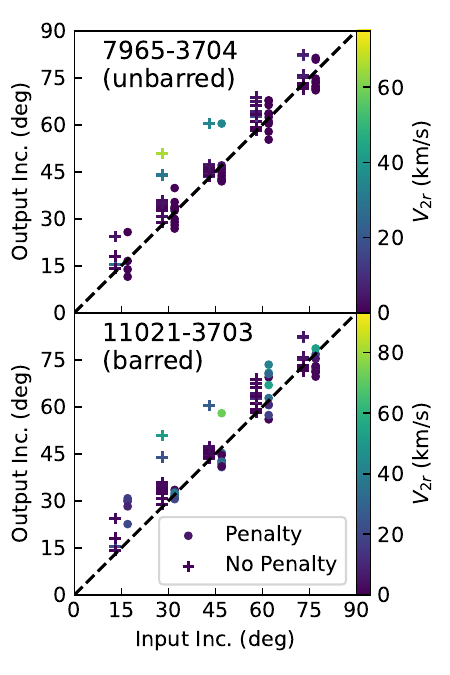}
    \caption{The effect of penalizing models that use high second-order velocities on the inclination bias. We construct a set of mock galaxies using the Nirvana velocity fields an unbarred galaxy (MaNGA plate and IFU number 7965-3704) with $V_{2t}$ and $V_{2r}$ close to zero on top, and a barred galaxy with elevated central $V_{2t}$ and $V_{2r}$ (11021-3703) on the bottom. We generate idealized models using these velocity profiles at different inclinations and add real residuals from galaxies with similar radius and inclination, creating a population of mock galaxies at varying inclinations. We then fit those mocks with Nirvana, allowing us to compare input and output parameters. We find that when unrestricted, Nirvana has a tendency to produce erroneously high inclination models (shown by the + symbols), which elevate $V_{2r}$ values due to degeneracies between inclination and $V_{2r}$ residuals. When we impose a penalty on high second-order velocity terms as described in Section \ref{sec:likelihood} (shown by the dots), the bias is greatly reduced and the spread between different residuals is tightened.}
    \label{penaltybias}
\end{figure}

This can be seen in Figure \ref{penaltybias}, which shows how well galaxy inclination is recovered during mock testing. We construct a set of mock galaxies by feeding model parameters derived from real Nirvana galaxy models at similar inclinations (one unbarred disk 7965-3704 and one barred disk 11021-3703 with a second-order velocity profile peaking at $\sim$50 km/s) and superimpose residuals from Nirvana models of comparable galaxies at varying inclinations to create sets of mock observations of the same galaxy at a range of inclinations. We then use Nirvana to fit these mock galaxies to test its ability to recover input parameters in realistic data. The model shows a tendency to fit erroneously high inclinations by utilizing similarly erroneous $V_{2r}$ values, as shown by the + symbols in in Figure \ref{penaltybias}. 

Models are also affected by the $m \pm 1$ degeneracy inherent in the bisymmetric model, as mentioned in Section\ref{sec:bisym}. This degeneracy between mode $m$ and modes $m\pm 1$ was noted by \citet{schoenmaker97} and \citet{spekkens07}, and we noted instances of this degeneracy influencing our model results during Nirvana development. In the case where $V_{2t} = V_{2r}$, we can use the angle-sum identity to rewrite Equation \ref{model} as: 
\begin{equation}
    \frac{V_{\rm los} - V_{\rm sys}}{\sin i} = V_t \cos\theta - V_2 \cos(\theta - 2\phi_b).
\end{equation}
for $V_2 \equiv V_{2t} + V{2r}$. That is, the combination of the second-order components mimic a first-order tangential component that is phase-shifted by $2\phi_b$, commonly referred to as a position-angle warp. This makes it possible for the model to effectively trade between $V_t$ and $V_2$ and their relevant position angles, $\phi$ and $\phi_b$, allowing Nirvana to create galaxy models where second-order motions erroneously dominate over first-order tangential motions instead of the other way around.

Because overinflated $V_{2t}$ and $V_{2r}$ values cause these issues, we disincentivize their overuse by imposing a penalty on the likelihood for models that have second order velocity terms that are large in comparison to the first order velocity using the following term:
\begin{equation}
    P_2 = w_2 \left(\frac{\bar{V}_{2t} - \bar{V_t}}{\bar{V_t}} + \frac{\bar{V}_{2r} - \bar{V_t}}{\bar{V_t}} \right),
\end{equation}
where barred quantities represent the means of the respective velocity profiles. $w_2$ is a separate coefficient that we determined through mock testing should be set to $w_2 = 500$ to produce results that capture bisymmetric velocity distortions when they are present but do not over-inflate them when they are not present. With this correction present, recovery of inclination in mock galaxies is much more faithful, as shown by the dots in Figure \ref{penaltybias}. We see both a lower average inclination bias and a smaller spread in variation for mocks with different residuals.

\subsection{Example Results}\label{sec:example}

\begin{figure}
    \centering
    \includegraphics[width=.5 \textwidth]{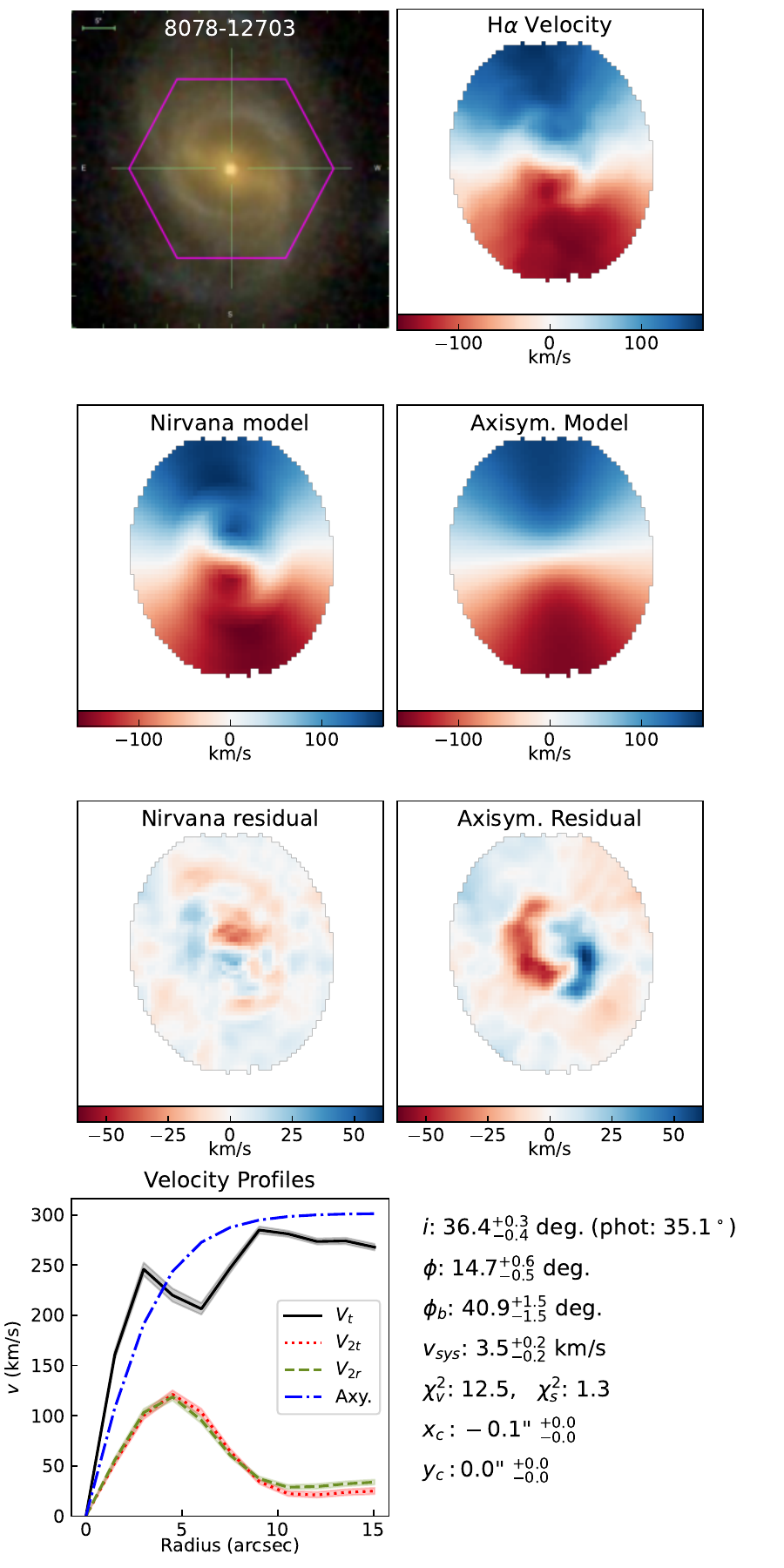}
    \caption{The Nirvana model of the gas-phase velocity field of barred MaNGA galaxy 8078-12703. Top row: the SDSS image of the galaxy with the MaNGA IFU boundary overlaid in magenta, and the gas-phase velocity field. Second row: The Nirvana velocity field model and the an axisymmetric using a parametric hyperbolic tangent rotation curve. Third row: Residuals for the above fits. Compared to the axisymmetric model, the residuals are significantly reduced and are much less spatially correlated, indicating a more suitable model. Bottom row: the best-fitting radial velocity profiles of the three velocity components fit by the Nirvana model ($V_t$ shown in solid black, $V_{2t}$ in dotted red, and $V_{2r}$ in dashed green) with $1\sigma$ errors, along with the rotation curve found by our parametric axisymmetric fitting algorithm (dot-dashed blue), and the rest of the parameters from the Nirvana model with $1\sigma$ errors.}
    \label{8078gasexample}
\end{figure}

\begin{figure}
    \centering
    \includegraphics[width=.5 \textwidth]{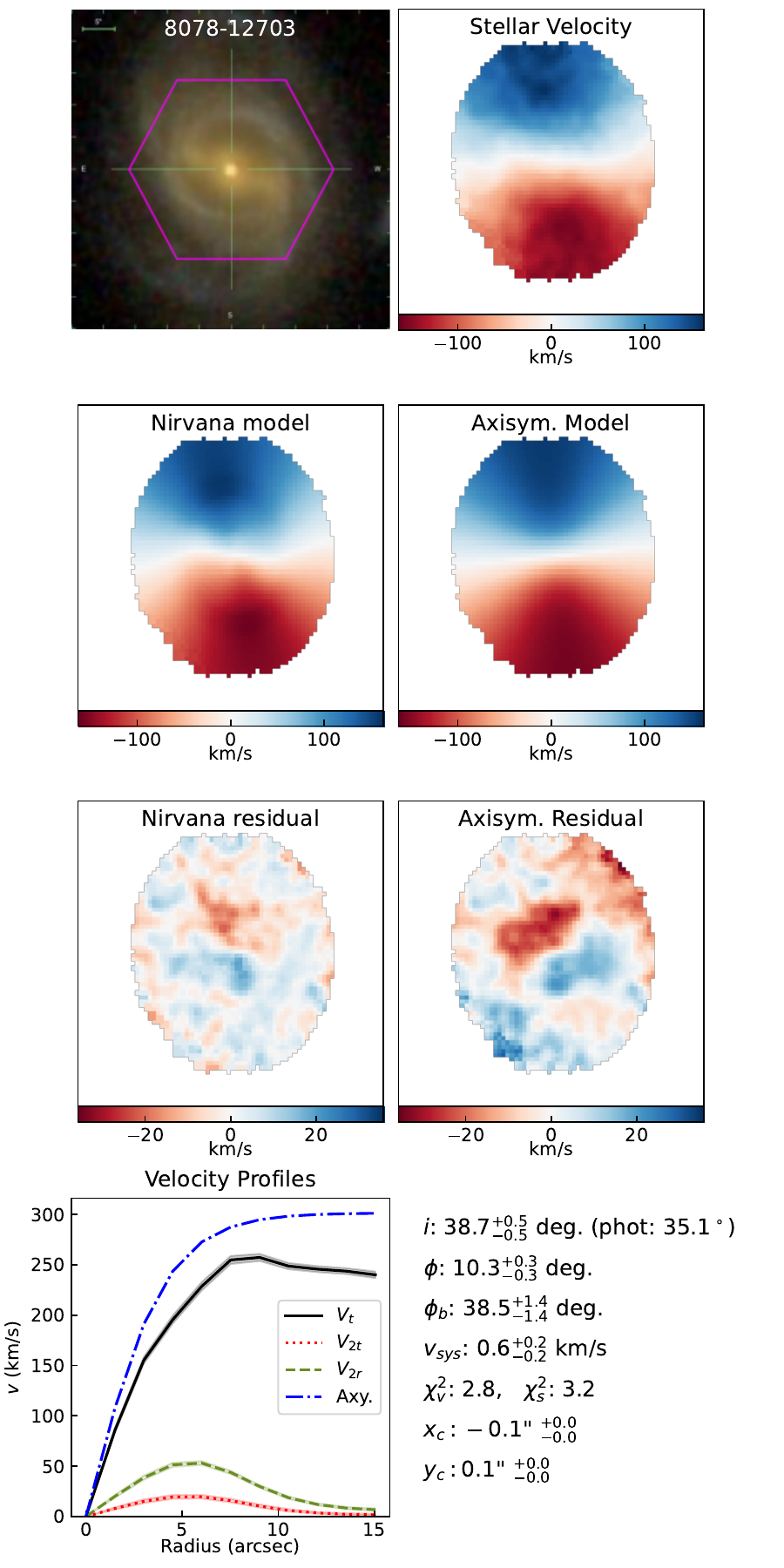}
    \caption{The same velocity field plots for 8078-12703 as Figure \ref{8078gasexample} but for the stellar velocity field. The magnitude of the velocity field disturbance caused by the bar is notably lower than for the gas-phase velocity field.}
    \label{8078starsexample}
\end{figure}

An example result from this model for barred MaNGA galaxy 8078-12703 is shown in Figures \ref{8078gasexample} and \ref{8078starsexample} for gas-phase and stellar velocity fields respectively. The non-axisymmetry of the bar is obvious in both the image and the velocity field, where a large central disturbance is visible in the otherwise orderly rotation of the disk. When the Nirvana model is applied, it recovers a first-order tangential rotation curve that roughly resembles a conventional model for a disk galaxy, rising quickly to a maximum value before leveling off at larger radii. The second-order components are present as a relatively large component of the rotation in the central part of the galaxy, but their influence quickly diminishes at larger radii as the influence of the bar lessens. 

\begin{figure}
    \centering
    \includegraphics[width=.5 \textwidth]{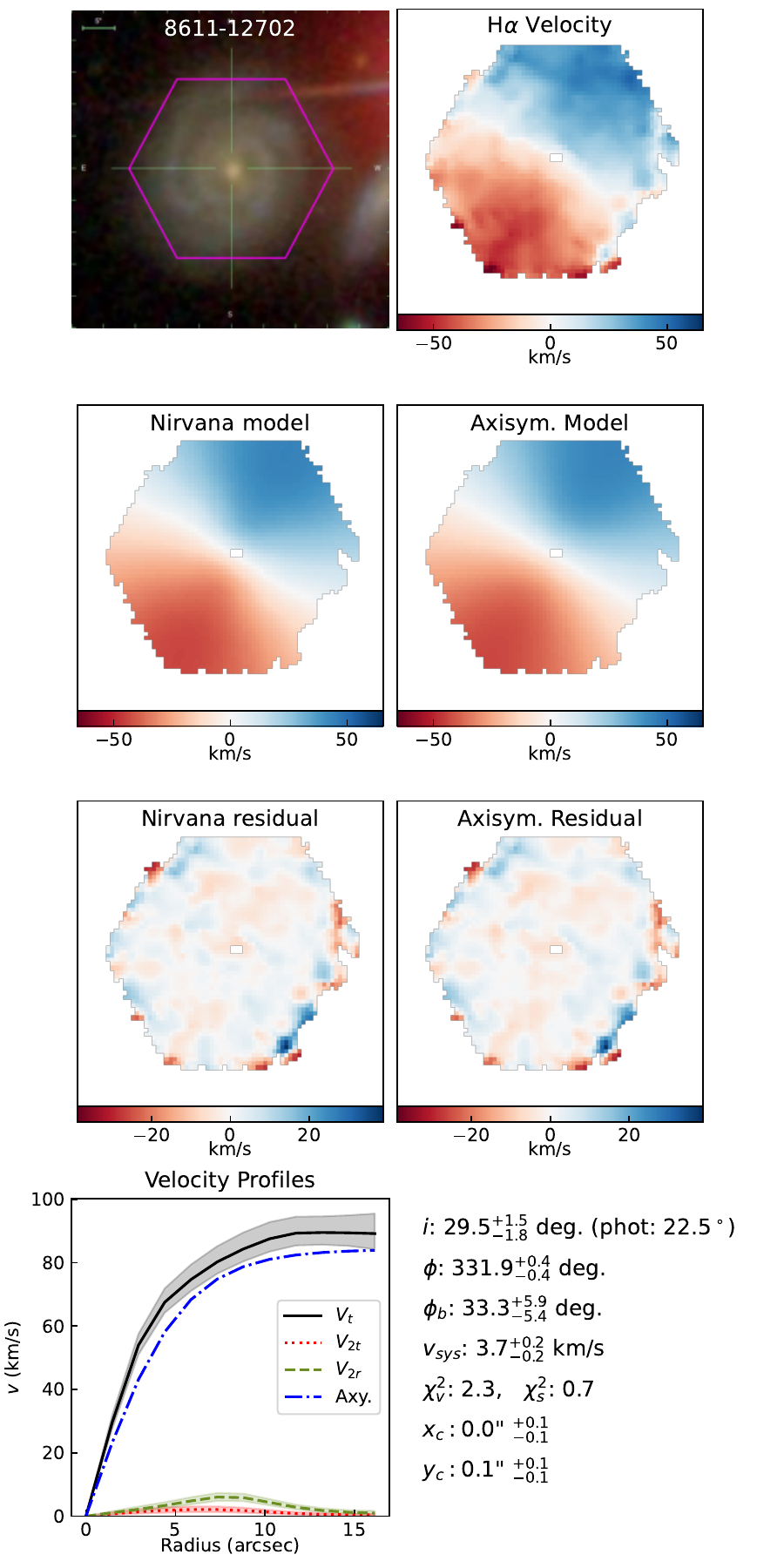}
    \caption{Velocity field plots for barred galaxy 8611-12702, a galaxy identified as barred by the GZ:3D volunteers but that does not display significant second-order velocity features in its velocity field. The axisymmetric model and the Nirvana model are both able to model the velocity field with similar rotation curves.}
    \label{8611example}
\end{figure}

\begin{figure}
    \centering
    \includegraphics[width=.5 \textwidth]{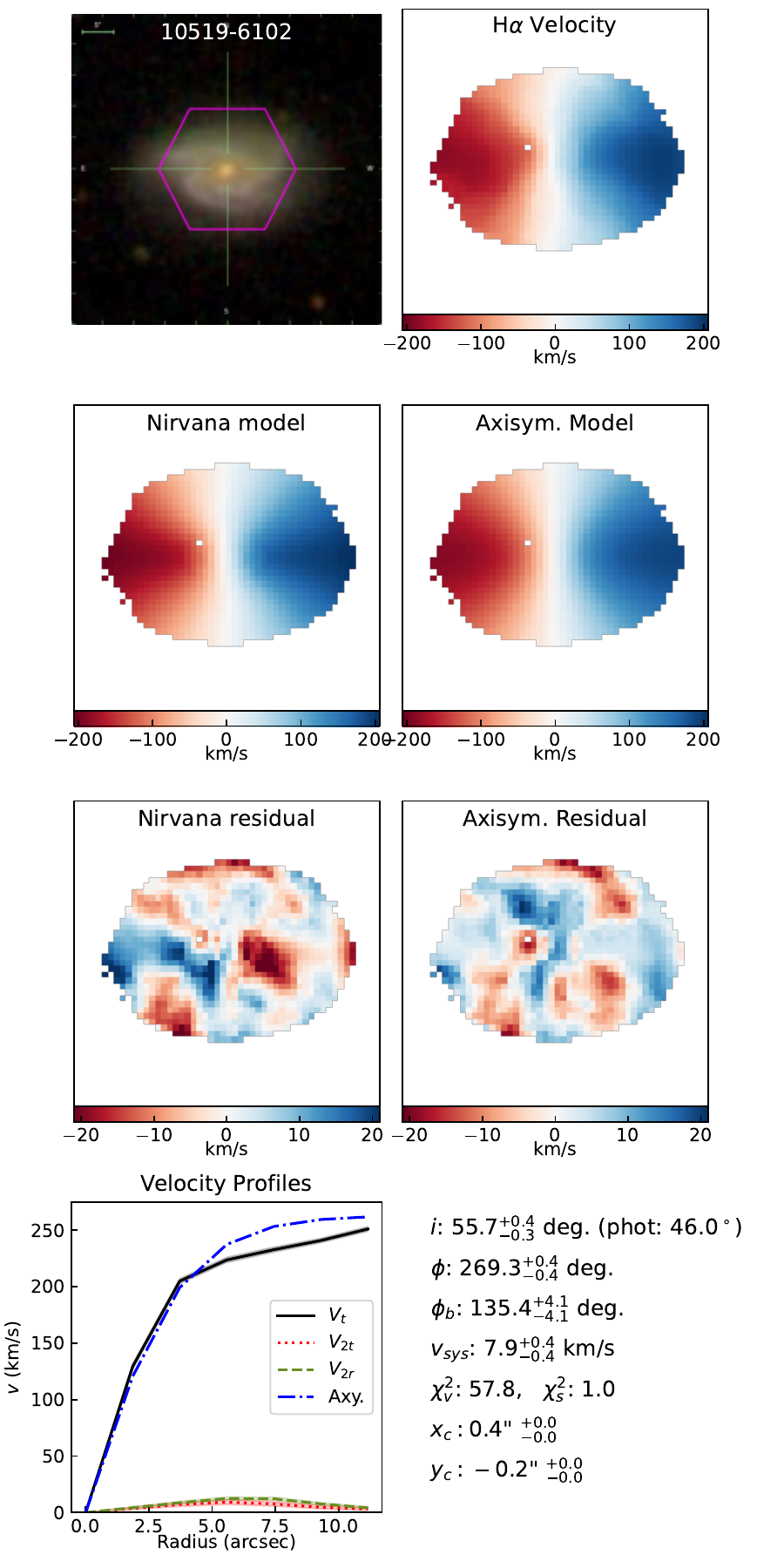}
    \caption{Velocity field plots for unbarred control galaxy 10519-6102. Like 8611-12702 (Figure \ref{8611example}), this galaxy does not have bisymmetric distortions and can be modeled well without significant contributions from second-order velocity terms.}
    \label{10519example}
\end{figure}

Figures \ref{8078gasexample} and \ref{8078starsexample} also demonstrate that, when compared to an axisymmetric model, Nirvana is able to more accurately model the observed 2D velocity field. The axisymmetric model leaves large and spatially-correlated residuals, indicating that the model is unable to capture all of the features seen in the data, whereas the Nirvana model's residuals are much smaller and much more randomly distributed. Figures \ref{8611example} and \ref{10519example} show Nirvana rotation curves for galaxies without strong bisymmetric velocity distortions (the first with a bar identified by GZ:3D and the second unbarred), producing results that are very similar to the hyperbolic tangent axisymmetric model. Further study is needed to determine why certain visually-identified bars do not have corresponding velocity field perturbations.

\begin{figure*}
    \centering
    \includegraphics[width=\textwidth]{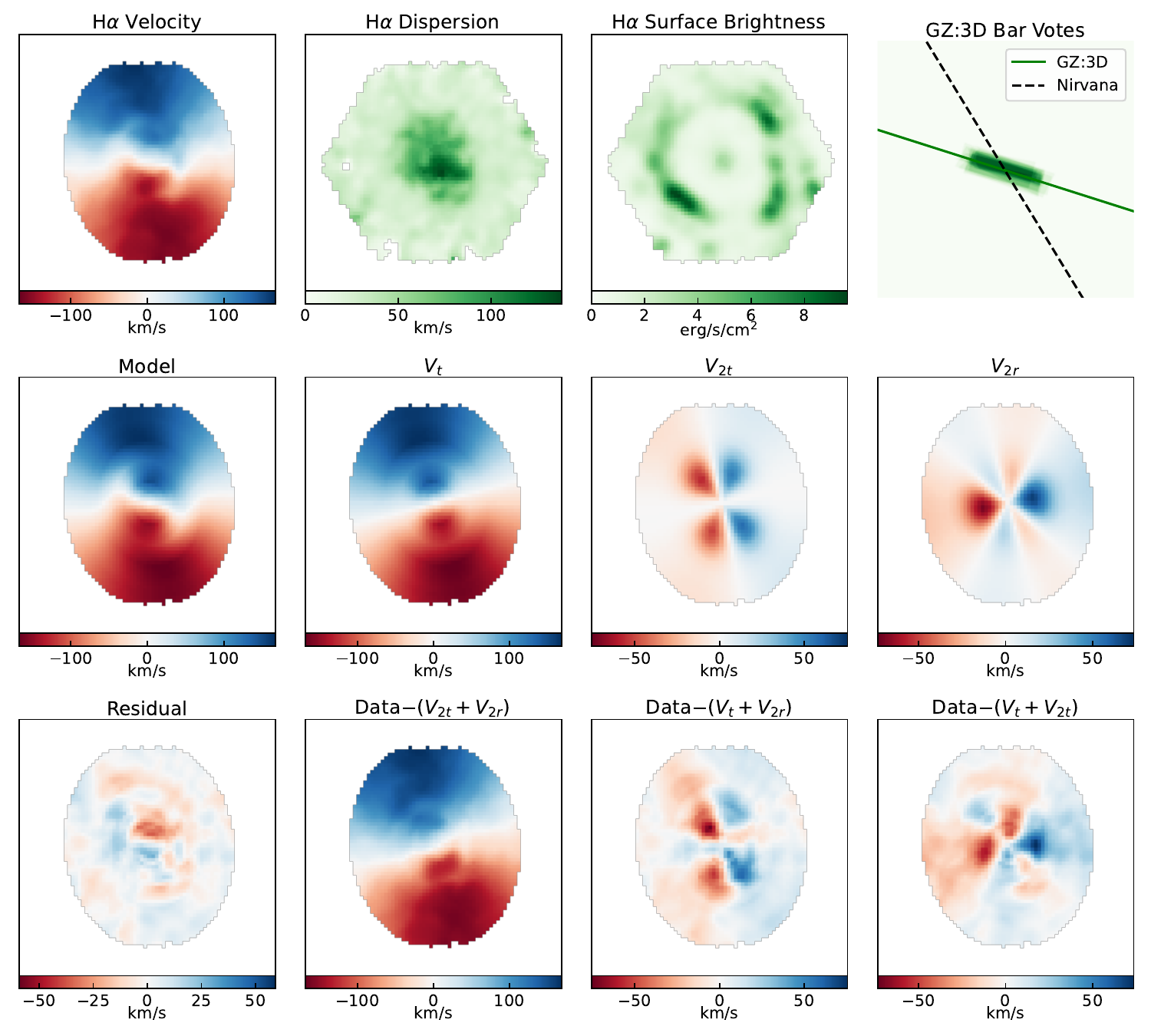}
    \caption{The separate pieces of MaNGA data that are fed to the Nirvana model and the individual components of the velocity field model for 8078-12703's gas phase velocity field (see Figure \ref{8078gasexample}). Top row: the MaNGA $H\alpha$ velocity field, velocity dispersion, and surface brightness. Top right: the bar classification votes from Galaxy Zoo: 3D and resulting on-sky bar position angles from GZ:3D and the independent Nirvana velocity model. Middle row: the Nirvana velocity field model, and all of the individual components of the model broken out separately. Bottom row: The residual of the velocity field model, and the component of the MaNGA velocity data that corresponds to the above velocity component.}
    \label{8078components}
\end{figure*}

The maps for the individual velocity modes of the Nirvana model as well as the components of the actual MaNGA data those modes are fitting can be seen in Figure \ref{8078components} for the gas-phase velocity field. The shapes of the components of the data generally match the shape of the velocity mode maps, justifying the physical premise of our model. The middle row of Figure \ref{8078components} shows a breakdown of the separate velocity components that make up the final velocity field model of the same galaxy. The bottom row shows the residuals left when subtracting different combinations of rotational terms from the original MaNGA data to leave only a single component in the data, yielding views of each component of the data that Nirvana is modelling. Comparing the velocity components of the second row to the residuals in the third row, we see close correspondence between our model terms and the noncircular motions present in the central bar region of the galaxy.

\section{Results} \label{sec:assessments}

In this section, we discuss the performance of the model on real and simulated galaxies in order to contextualize its results. 

\subsection{Projection biases} \label{sec:pabbias}

\begin{figure}
    \centering
    \includegraphics[width=.5 \textwidth]{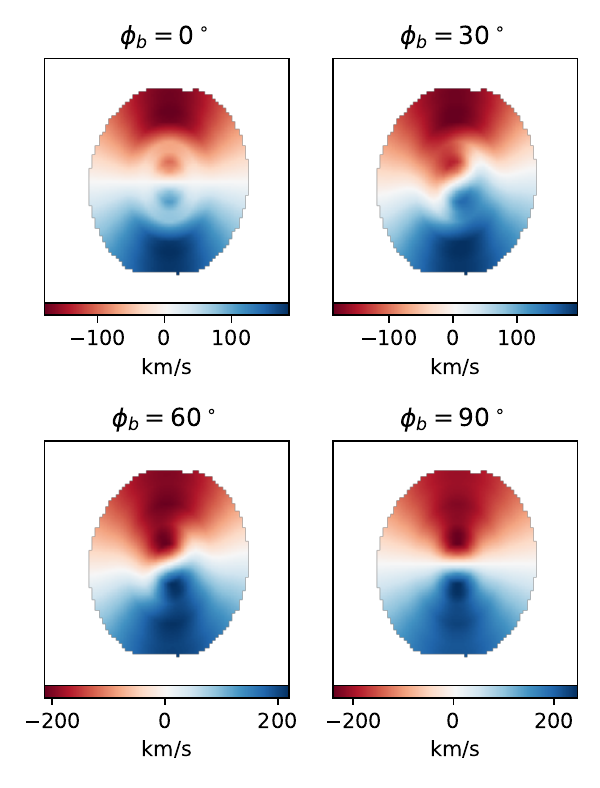}
    \caption{A comparison of different relative position angles between the dominant first-order and secondary bar component in model velocity fields. These mock galaxies are based off of the Nirvana rotation curves for MaNGA galaxy 8078-12703 with an inclination of $45^\circ$. For relative bar position angles that are diagonal or diagonal in-plane, the bisymmetric motion creates clear distortions in the shapes of the isovelocity contours, allowing Nirvana to recognize the bisymmetric velocity component. However, for bars aligned or anti-aligned with the major or minor axis (in-plane angular difference of $0^\circ$ or $90^\circ$), the isovelocity contours only change in magnitude rather than shape, an effect that can be modeled without a bisymmetric component.}
    \label{relpabcomparison}
\end{figure}

When modeling bisymmetric distortions in velocity fields caused by bar in disk galaxies, the angular difference between the position angles of the major axis $\phi$ (the first order velocity component) and the bar $\phi_b$ (the second order velocity component) greatly affects how the bar appears in the line-of-sight velocity data. Bars that are diagonal to the major axis will create obvious distortions in the velocity field, whereas bars that are aligned or anti-aligned with the major or minor axis will only appear as small fluctuations in the dominant first order rotational component, as shown in Figure \ref{relpabcomparison}. Nirvana often models these disturbances without second-order velocity components, leading to significant difficulties in accurately recovering aligned and anti-aligned bars.{ This is similar to DiskFit, which also systematically underestimates second-order motions for aligned and anti-aligned bars as compared to diagonal bars \citep{spekkens07}. \citet{randri16} finds that second-order velocities become proportionally much smaller than second-order photometric amplitudes, which mirrors our results.}

\begin{figure}
    \centering
    \includegraphics[width=.5 \textwidth]{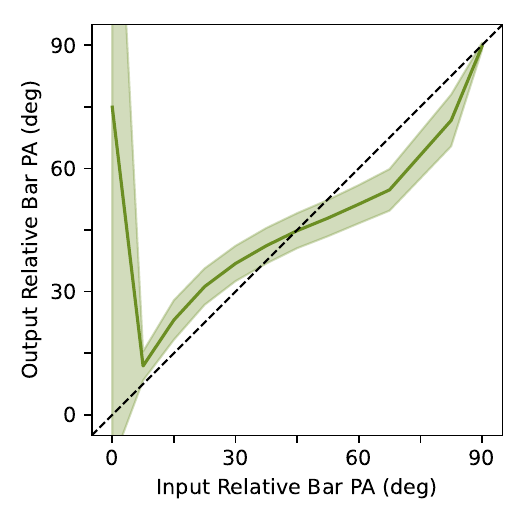}
    \caption{The recovered relative position angles $\phi_b$ and errors on posteriors from a set of mock galaxies similar to those shown in Figure \ref{relpabcomparison} projected onto the plane of the sky. Relative position angles that are roughly $45^\circ$ are recovered faithfully, but diagonal bars are always biased towards $45^\circ$, sometimes leading to biases over $5-10^\circ$. Aligned and anti-aligned bars are difficult to distinguish in velocity data, leading to inflated or unrealistic errors on bisymmetric position angle due to a lack of constraint on the model (as seen in the case of the aligned bar in this plot), but they have no inherent bias.}
    \label{pabbias}
\end{figure}

In the set of mocks shown in Figure \ref{pabbias}, we see that galaxies with $\phi_b$ values that are close to aligned/anti-aligned, Nirvana has a preference for increasing relative $\phi_b$ values between $0^\circ$ and $45^\circ$ and decreasing values between $45^\circ$ and $90^\circ$. The effect of this is to bias $\phi_b$ to be closer to a $45^\circ$ or $135^\circ$ offset from $\phi$ than reality, and the second-order velocity profiles for these biased bars are often less than the input velocity profiles. 

\begin{figure}
   \centering
    \includegraphics[width=.5 \textwidth]{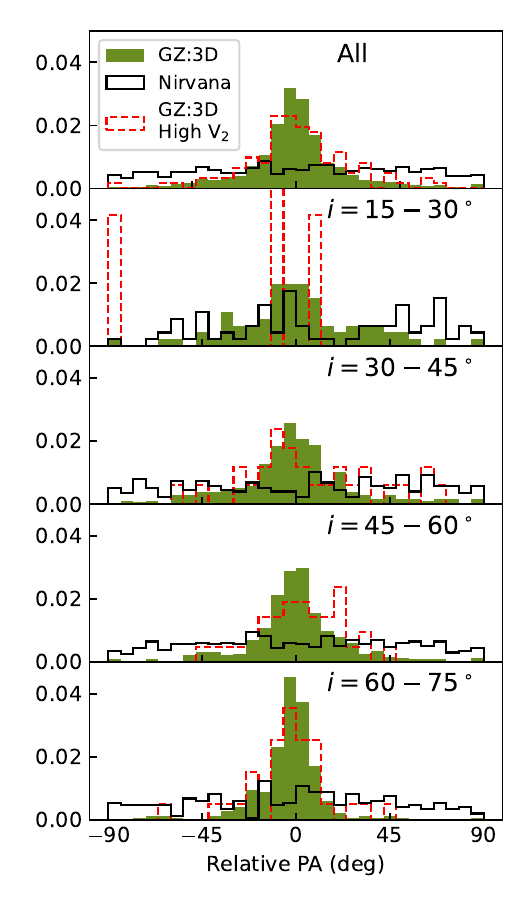}
    \caption{Histograms showing the distribution of on-sky relative position angles recovered by Nirvana and Galaxy Zoo:3D for the entire sample of barred galaxies (top) as well as broken down into inclination bins. Nirvana inherently biases towards bars that are at a $45^\circ$ angle to the major axis because those bars cause larger kinematic asymmetries, but that small bias is overwhelmed by the large GZ:3D bias towards bars that are aligned with the major axis. This bias arises because they are not as distorted by projection effects and are thus easier for volunteers to identify, a problem that worsens with inclination. {The subsample of the 10\% of Nirvana galaxies with the highest $V_2$ terms (red dashed histograms) track more closely to the expected distribution of bar angles, validating the accuracy of the model for these galaxies.}}
    \label{allrelpabhists}
\end{figure}

The origins of this bias are unclear. Because more diagonal $\phi_b$ produces a stronger bisymmetric distortion than a more aligned one, Nirvana requires smaller second order velocity components to explain the same bisymmetric features in the velocity field. This minimizes the $P_2$ penalty in the likelihood necessary for inclination and second-order magnitude corrections (see Section \ref{sec:likelihood}), yielding a potentially more favorable outcome. However, when $P_2$ is turned off in the code, the bias still remains so this cannot be the explanation. 

\subsection{Comparison with imaging}
\label{sec:gz3d}

\begin{figure}
    \centering
    \includegraphics[width=.5 \textwidth]{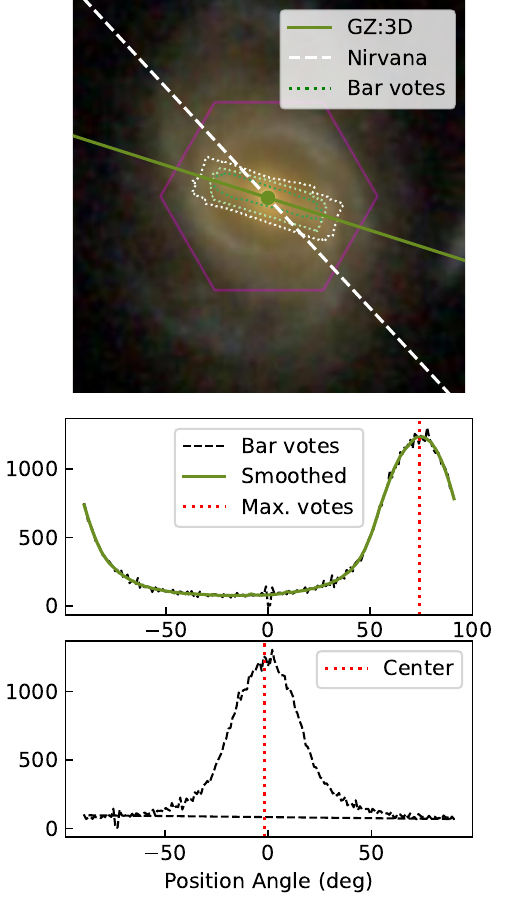}
    \caption{A set of subplots summarizing the method used to distill the GZ:3D bar classifications down to a single position angle for a galaxy. Top: The SDSS image of MaNGA galaxy 8078-12703 overlaid with the extent of the MaNGA IFU (magenta), the fraction of votes indicating the presence of a bar (dotted contours), the bisymmetric position angle from the Nirvana model (white dashed) and the GZ:3D bar position angle derived using this method (solid green). The weighted center of the bar votes is marked as a green circle. Middle: The number of GZ:3D bar votes from volunteers that fall into different azimuthal bins (black dashed) are smoothed to remove high-frequency noise (green) and the peak number of smoothed votes is used as a first approximation for the bar position angle (red dotted). Bottom: The azimuthal slices are recentered on this approximation (black dashed) and the weighted center of the peak is calculated (red dotted) to reduce the effect of asymmetric or bimodal peaks. This final position angle is used as the bar position angle in the top subplot. More examples can be seen in Figure \ref{imagemosaic}.}
    \label{gzbarang}
\end{figure}

In order to validate Nirvana's bar position angles, we compare our results to those of GZ:3D \citep[][see Section \ref{sec:sample}]{masters21}. Because GZ:3D treats each pixel individually, the GZ:3D bars are irregular in shape, making it difficult to define a bar position angle. We developed the following procedure (shown in Figure \ref{gzbarang}) for finding a representative bar position angle for each galaxy. First, we use the votes as weights to find the weighted center of the bar mask, which we take to be the center of the bar. Next, we divide the image into on-sky azimuthal bins, adding up the bar votes within each bin to create an azimuthal distribution of bar votes. We then use a Savitzky-Golay smoothing filter to remove higher order noise from this distribution to obtain a more continuous curve. We then adjust the distribution so its maximum is in the center, yielding a smooth and approximately symmetrical distribution of bar votes. Finally, we calculate the weighted mean of the whole distribution, which gives us our final bar position angle that is robust to visual inspection and relatively resistant to irregular bar shapes and volunteer misclassifications. The process is summarized in Figure \ref{gzbarang}.

{Though Nirvana relative position angles are roughly evenly distributed, (Figure \ref{allrelpabhists}),} the GZ:3D volunteer classifications themselves display a bias towards bars aligned with the major axis that is present in the final GZ:3D data set. Projection effects lead to a nonuniform distortion in azimuthal angles in high inclination galaxies, meaning that even a uniform distribution of on-sky bar angles will become biased towards major axis bars when transformed to in-plane coordinates. In addition, because bars along the minor axis are foreshortened due to projection effects, they can be difficult to distinguish from a bulge in inclined galaxies \citep{bureau1999, binney08}, leading to a likely underreporting of bars close to the minor axis by GZ:3D volunteers. These confounding factors lead to a significant overrepresentation of bars that are closely aligned with the major axis in the GZ:3D sample, which in turn introduces the same bias into the Nirvana-MaNGA sample. Thus, we find a drastic dearth of bars perpendicular to the major axis, especially at higher inclinations where projection effects are larger. This is seen in the solid green histograms in Figure \ref{allrelpabhists}. 

{We find different results for galaxies with large second-order velocity components in Nirvana models. We define a subsample consisting of the 10\% of Nirvana-MaNGA barred galaxies with the highest gas-phase $V_{2r}$ values at 1/3 of their radius ($V_{2r} \gtrsim 50$ km/s). We choose this characteristic for constructing the subsample because 1) bars are associated with radial motions; and 2) the influence of bars greatly diminishes beyond corotation \citep{binney08}, so we focus on the inner region of the galaxy. These galaxies, shown as the dashed red histograms in Figure \ref{allrelpabhists}, follow the expected distribution of bar angles more closely. This indicates that random errors seem to affect bar measurements for galaxies with strong second-order components less than those with weak second-order components, producing results that more closely match imaging data.}

\begin{figure}
    \centering
    \includegraphics[width=.5 \textwidth]{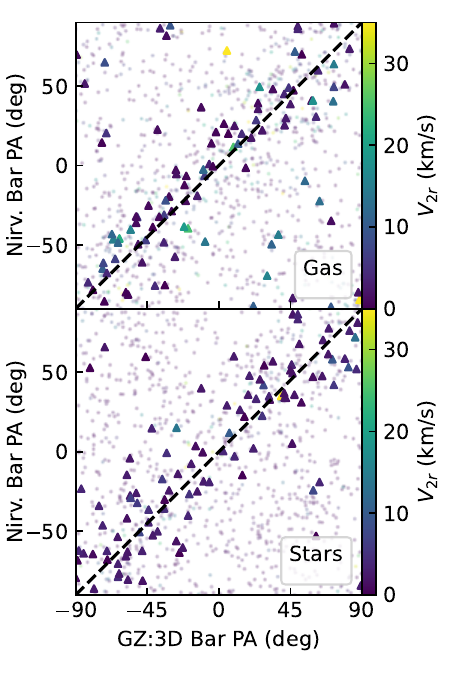}
    \caption{Comparisons between the bar position angles derived from Galaxy Zoo: 3D and the on-sky bisymmetric kinematic position angles derived from Nirvana for gas-phase (top) and stellar (bottom) velocity fields for barred galaxies in MaNGA. Our subsample of galaxies in top 10\% of $V_{2r}$  magnitude ($\gtrsim 50$ km/s) at 1/3 of their radius (triangles) show a strong correspondence between kinematically-derived position angles for bisymmetric terms in Nirvana and the imaging-derived bar position angles from GZ:3D, while the Nirvana-MaNGA sample as a whole (circles) shows a weaker correspondence. This indicates that when Nirvana recovers significant second-order motions in a galaxy, it tends to agree with visual classifications on bar angle, although the correspondence is tighter for gas-phase velocity fields than for stellar velocity fields.}
    \label{pab1to1}
\end{figure}

{This pattern is also seen when comparing bar angles directly.} We find a little correspondence between the bar position angles between GZ:3D and the Nirvana-MaNGA barred sample overall{, which may be understood in the framework of \citet{randri16} as a misalignment between the photometric Fourier components of the bar and the second-order velocity components from the rotation model}. However, the correspondence is greater for galaxies with more bisymmetric motion{, as defined using the subsample from above.} Galaxies in this subsample display a much tighter correspondence with GZ:3D in bar position angle, and the remainder of the galaxies with comparatively small second-order motions show little correlation, as shown in Figure \ref{pab1to1}. Thus, we find that only a fraction of visually-identified galactic bars are accompanied by strong non-circular motions according to Nirvana. Further study is needed to determine the source of the discrepancy between visual and kinematic bars. 

\begin{figure*}
    \centering
    \includegraphics[width=\textwidth]{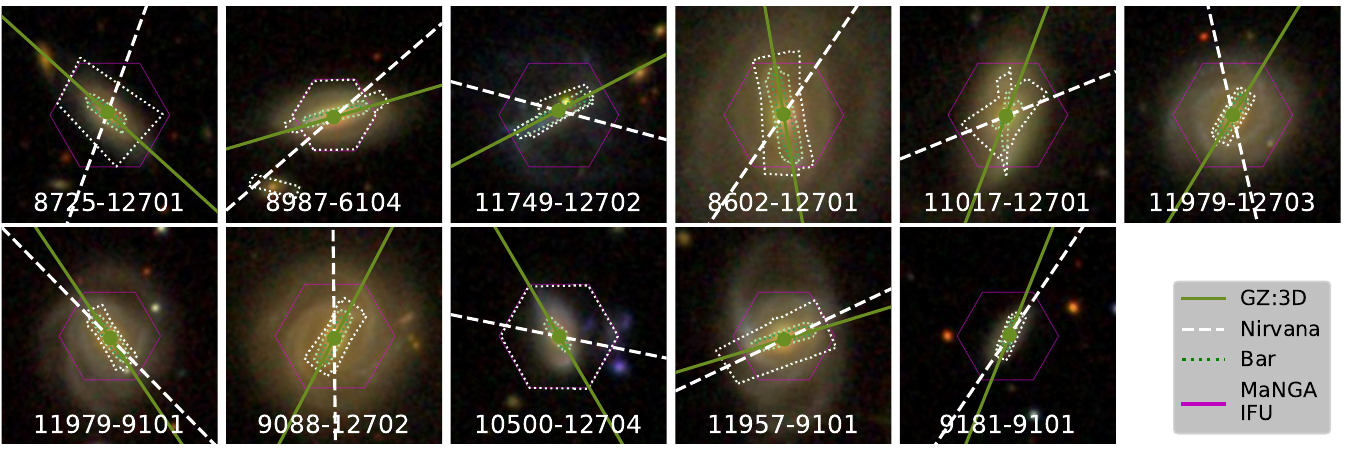}
    \caption{A random selection of SDSS images of Nirvana-MaNGA galaxies from the subsample with the highest $V_{2r}$ magnitudes. Overlaid are the boundaries of the MaNGA IFU (magenta), the GZ:3D bar position angle (solid green), the Nirvana bisymmetric position angle (dashed white), the GZ:3D bar votes (dotted contours), and the MaNGA plate and IFU identifiers. Some galaxies show a tight correspondence between the visually-identified GZ bar and the kinematically-identified Nirvana bar, while others show a large difference.}
    \label{imagemosaic}
\end{figure*}

Several visual examples of GZ:3D/Nirvana bar correspondence within the high-$V_{2r}$ subsample are found in Figure \ref{imagemosaic}.

\subsection{Velocity components} \label{sec:recovery}

\begin{figure}
    \centering
    \includegraphics[width=.5 \textwidth]{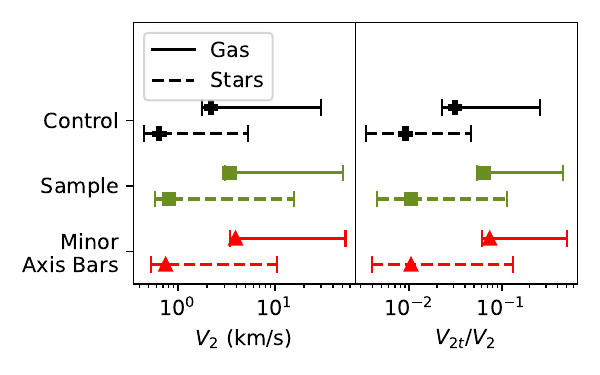}
    \caption{The distributions of the magnitudes of second-order radial velocity profiles $V_2$ at 1/3 the radius of the Nirvana models (left) and the ratio between the $V_2$ and $V_t$ at that radius (right) for both the Nirvana-MaNGA barred sample (green squares) and the control sample (black pluses). Medians and 68\% intervals are marked for both gas-phase (solid lines) and stellar (dashed lines) velocity field models. There are significant differences in radial motions for both gas and stellar velocity fields, indicating that bars are indeed associated with non-circular motions, but the magnitude of the motions is much greater for gas than for stars. We also find that bars that are aligned with the minor axis (red triangles) differ only slightly from other bars, indicating that Nirvana's bias is minimal.}
    \label{v2rerrbars}
\end{figure}

Nirvana finds higher average second-order velocity components in the sample of barred galaxies than in the controlled sample of unbarred galaxies, confirming that bars are indeed associated with elevated second-order motions in some galaxies. This trend can be seen in Figure \ref{v2rerrbars}. The median $V_{2}$ magnitude measured at 1/3 of the Nirvana model's radius (the approximate peak of bar velocity profiles, from inspection) is significantly higher in the gas-phase velocity fields of barred galaxies, with the upper tail of the distribution extending significantly higher indicating a greater fraction of galaxies with larger non-circular motions. The difference is also present in the stellar velocity fields but the difference is not as large, and the magnitudes of second-order motions is not as high overall, indicating that bars have a lesser influence on stellar kinematics than gas kinematics. We find only a slight difference in $V_{2}$ magnitude among galaxies with bars close to the minor axis in gas-phase velocity fields and little discernible difference in stellar velocity fields, confirming that Nirvana has little significant velocity bias for aligned or diagonal bars.

\begin{figure*}
    \centering
    \includegraphics[width=\textwidth]{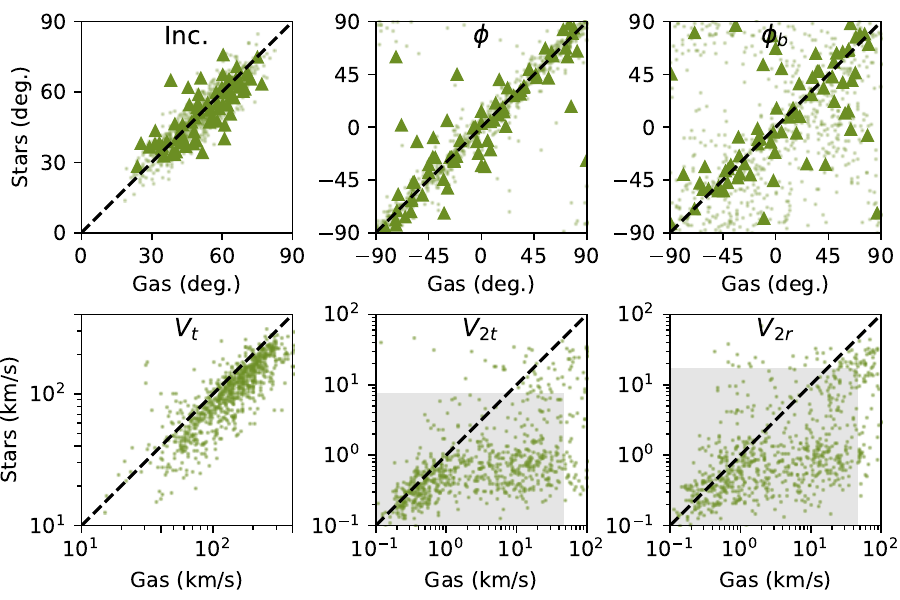}
    \caption{Comparisons between the Nirvana output for stellar- and gas-phase velocity field model parameters for the Nirvana galaxies where the model finished running for both velocity fields. The top row contains global galaxy parameters, showing tight agreement in inclination (left) and first-order position angle $\phi$ (middle). {The relative position angle $\phi_b$ (right) shows moderate agreement, but has much tighter agreement for the subsample of galaxies with high $V_{2r}$ (triangles), as outlined in the text.} The bottom row compares velocity profile magnitude at 1/3 the model radius, showing good agreement for $V_t$ and smaller $V_{2t}$ and $V_{2r}$ values in most stellar models. {Median errors on velocities are $\sim$ 3-4 km/s. The top and right boundaries of the shaded regions denote the 90th percentile of that velocity component, so the galaxies outside the shaded region are considered to have large $V_2t$ and/or $V_2r$ values.}}
    \label{gasstarcomparisons}
\end{figure*}

Overall, the Nirvana models for stellar- and gas-phase velocity fields agree well on global galaxy parameters like inclination and the first-order position angle for galaxies where both model runs finished. {First-order tangential rotation speeds track closely, though stellar speeds are lower than gas speeds due to asymmetric drift \citep{binney08}, an effect warranting further exploration using this data. Gas- and stellar-phase models diverge more with the second-order velocity components, as seen in Figure \ref{gasstarcomparisons}. As with the comparison with GZ:3D bar position angles, there is more agreement between stellar- and gas-phase relative position angles when using the subsample of galaxies with large non-circular motions, indicating greater consistency when the model is more constrained. We also see systemic evidence of lower $V_2$ magnitudes at 1/3 of the model radius in comparison to $V_t$ in stellar velocity fields than in corresponding gas velocity fields. This bolsters earlier conclusions from inspection that the second-order components are less prominent overall in stellar velocity fields than in their gas counterparts, though galaxies with higher $V_{2r}$ values display less mismatch.} Further study is needed to investigate the differences between the population of barred galaxies with stellar $V_2$ values that hew close to their gas $V_2$ and those that do not.

{When examining stellar- and gas-phase velocity components, it is apparent that Nirvana sometimes recovers unphysically large rotational speeds. Upon inspection, many of these galaxies either have a kinematic center that is greatly misaligned with the center of the MaNGA IFU or are out of kinematic equilibrium due to some recent perturbation. Though effort was made to remove actively merging galaxies from the Nirvana sample (see Section \ref{sec:sample}), our method may not detect actively merging galaxies that do not have two visible nuclei in SDSS images, so some kinematically disrupted galaxies are still included in the sample. Such galaxies are likely not well described by a rotating thin disk, so their modeled rotational speeds cannot be reliably compared to other more regular rotating galaxies. When working with Nirvana data, it is important to assess the credibility of any severely outlying results to ensure the model was working within its intended use case as uncertainties may underestimate total errors, and further validation may be necessary to ensure reliability when applying the code to non-MaNGA data.}

\section{Summary}
\label{sec:summary}

The Nirvana software package is a Bayesian velocity field modeling code that can reliably fit both circular and bisymmetric motions in 2D kinematic data for spiral galaxies. {We build on previous works (e.g. DiskFit, \citealt{spekkens07,sellwood15}; \texttt{XookSuut}, \citealt{lopezcoba21}; \textsc{2DBAT}, \citealt{oh18}), adding further capabilities for lower-spatial-resolution kinematic data like modeling velocity dispersion profiles and PSF convolution, and we use a Bayesian framework with physically-informed priors to improve the reliability of our results. The result is a code that is suitable to run on optical IFU data of galaxies that are fewer than 10 beam widths across, surpassing capabilities of previous codes.}

We construct our Nirvana-MaNGA sample of over 1000 barred galaxies using the volunteer classifications of barred galaxies from the GalaxyZoo: 3D catalog, along with a control sample of MaNGA disk galaxies matched to the main sample in color, mass, effective radius, and axis ratio. The Nirvana model has been tested against real and mock data to produce reasonable and physically-motivated velocity field models for stellar and gas-phase kinematics in a wide variety of spiral galaxies by using custom prior and likelihood functions and sanitizing its own input data. The resultant models have only relatively small  biases in inclination and bar position angle that we explore above.

We find that a significant fraction of visually-identified bars do not have discernible higher-order terms in their velocity fields, a conclusion meriting further study Nirvana's on-sky second-order position angles show a correspondence with imaging-based bar angles from GZ:3D despite notable biases from projection effects, confirming a relationship between visually-identified bisymmetric structures and kinematic disturbances from non-circular motions. We also find that Nirvana reliably recovers more second-order velocity modes in barred galaxies than in unbarred galaxies, validating the dynamical properties of bars in the largest sample of real galaxies yet assembled. Nirvana finds significantly higher second-order velocity modes in gas-phase velocity fields than in stellar velocity fields and finds no non-circular terms in many galaxies that would be visually classified as barred, warranting further investigation into the effects of bars on different kinematic components in galaxy centers. Our sample of non-parametric second order rotation curves will also allow for the design of an empirically-motivated parametric velocity field model of higher order motions in barred galaxies, which would improve the speed and usefulness of these models.

Our spaxel-by-spaxel maps of non-circular motion magnitudes in MaNGA barred spirals allow further study of the influence of bars on other galaxy properties. It is possible to directly search for a correlation between elevated non-circular motions within bars and radial-mixing-driven flattening of stellar population gradients and other population differences in barred galaxies, as has been seen with existing visually-identified barred galaxy samples \citep[e.g.][]{fraser19, fraser20, krishnarao20}. Our physically-motivated measures of non-circular motions may also provide a new perspective on the influence of kinematic asymmetry on Tully-Fisher scatter \citep{bloom17, andersen2013}, provide new methods for finding galactic inflows and outflows, allow for new estimations of asymmetries in dark matter halos \citep{sellwood10}.

The Nirvana code can also easily be applied to other data sets as long as they have information on kinematics, surface brightness, and PSF. The Nirvana-MaNGA sample provides a comprehensive baseline of the kinematic properties of barred galaxies in the local Universe, so a sample of Nirvana models of more distant galaxies would allow for the study of the evolution of bar kinematics over the course of galactic evolution.

\acknowledgements

We acknowledge Andrew Leung, whose early work on a similar topic assisted the beginning of this project.

We acknowledge use of the lux supercomputer at UC Santa Cruz, funded by NSF MRI grant AST 1828315.

Funding for the Sloan Digital Sky Survey IV has been provided by the
Alfred P. Sloan Foundation, the U.S. Department of Energy Office of
Science, and the Participating Institutions. SDSS-IV acknowledges
support and resources from the Center for High-Performance Computing at
the University of Utah. The SDSS web site is www.sdss.org.

SDSS-IV is managed by the Astrophysical Research Consortium for the
Participating Institutions of the SDSS Collaboration including the
Brazilian Participation Group, the Carnegie Institution for Science,
Carnegie Mellon University, the Chilean Participation Group, the French
Participation Group, Harvard-Smithsonian Center for Astrophysics,
Instituto de Astrof\'isica de Canarias, The Johns Hopkins University,
Kavli Institute for the Physics and Mathematics of the Universe (IPMU) /
University of Tokyo, Lawrence Berkeley National Laboratory, Leibniz
Institut f\"ur Astrophysik Potsdam (AIP),  Max-Planck-Institut f\"ur
Astronomie (MPIA Heidelberg), Max-Planck-Institut f\"ur Astrophysik (MPA
Garching), Max-Planck-Institut f\"ur Extraterrestrische Physik (MPE),
National Astronomical Observatories of China, New Mexico State
University, New York University, University of Notre Dame,
Observat\'ario Nacional / MCTI, The Ohio State University, Pennsylvania
State University, Shanghai Astronomical Observatory, United Kingdom
Participation Group, Universidad Nacional Aut\'onoma de M\'exico,
University of Arizona, University of Colorado Boulder, University of
Oxford, University of Portsmouth, University of Utah, University of
Virginia, University of Washington, University of Wisconsin, Vanderbilt
University, and Yale University. 

\software{Astropy \citep{2013A&A...558A..33A, 2018AJ....156..123A};
Numpy \citep{harris2020array};
Scipy \citep{2020SciPy-NMeth}; matplotlib \citep{Hunter:2007}; fftw \citep{fftw}}

\bibliography{ref}

\end{document}